\documentclass[aps,prb,twocolumn,showpacs,a4paper]{revtex4}

\usepackage{graphicx}
\usepackage{amssymb}%
\usepackage{epsfig}
\usepackage{amsmath}
\usepackage{subeqnarray}
\usepackage{ulem}
\usepackage{color}
\usepackage{stmaryrd}
\usepackage{multirow}
\usepackage{subfigure}
\usepackage[pdftex,breaklinks=true,colorlinks=true]{hyperref}
\usepackage[active]{srcltx}
\newcommand{\qzero}{$\mathbf q=0$ }
\newcommand{\sqrtsqrt}{$\sqrt3\times \sqrt3$ }

\newcommand{\journal}[4]
{\ifthenelse{\equal{#1}{pr}}{
Phys Rev. {\bf #2}, \href{http://link.aps.org/abstract/PR/v#2/e#3}{#3} (#4)}
{\ifthenelse{\equal{#1}{prl}}{
\prl {\bf #2}, \href{http://link.aps.org/abstract/PRL/v#2/e#3}{#3} (#4)}
{\ifthenelse{\equal{#1}{prb}}{
\prb {\bf #2}, \href{http://link.aps.org/abstract/PRB/v#2/e#3}{#3} (#4)}
{\ifthenelse{\equal{#1}{pra}}{
\pra {\bf #2}, \href{http://link.aps.org/abstract/PRA/v#2/e#3}{#3} (#4)}
{\ifthenelse{\equal{#1}{arxiv}}{preprint
\href{http://arxiv.org/abs/#2.#3}{arXiv:#2.#3}}
{\ifthenelse{\equal{#1}{rmp}}{
\rmp {\bf #2}, \href{http://link.aps.org/abstract/RMP/v#2/e#3}{#3} (#4)}
{\ifthenelse{\equal{#1}{cond-mat}}{preprint
\href{http://arxiv.org/abs/cond-mat/#2}{cond-mat/#2}}
{\ifthenelse{\equal{#1}{pre}}{
\pre {\bf #2}, \href{http://link.aps.org/abstract/PRE/v#2/e#3}{#3} (#4)}
{#1 {\bf #2}, #3 (#4)}}}}}}}}}

\begin{document}

\title{The kagome antiferromagnet: a  chiral  topological spin liquid?}
\author{Laura Messio,$^1$ Bernard Bernu$^2$ and Claire Lhuillier$^2$}
\affiliation{
$^1$ Institute of Theoretical Physics, Ecole Polytechnique F\'ed\'erale de Lausanne (EPFL), CH-1015 Lausanne, Switzerland.\\
$^2$ Laboratoire de Physique Th\'eorique de la Mati\`ere Condens\'ee, UMR 7600 CNRS, Universit\'e Pierre et Marie Curie,  75252 Paris Cedex 05, France.
}

\begin{abstract}
Inspired by the recent discovery of  a new instability towards a chiral phase  of the classical Heisenberg model on the kagome lattice, we propose a specific chiral spin liquid that reconciles different, well-established results concerning both the classical and quantum models.
This proposal is analyzed in  an extended mean-field  Schwinger boson framework encompassing time reversal symmetry breaking phases which allows both a classical and a quantum phase description.
At low temperatures, we find quantum fluctuations favor this chiral phase, which is stable against small perturbations of second and third neighbor interactions. For spin-1/2  this phase may be,  beyond mean-field, a chiral gapped spin liquid.  Such a phase is consistent with Density Matrix Renormalization Group results of  Yan et al. (Science \textbf{322}, 1173 (2011)). Mysterious features of the low lying excitations of exact diagonalization spectra also find  an explanation in this framework.
Moreover, thermal fluctuations compete with quantum ones and induce a transition from this flux phase to a planar zero flux phase at a non zero value of the renormalized  temperature ($T/\mathcal S^2$),  reconciling these results with those obtained for the classical system.
\end{abstract}
\pacs{75.50.Ee,75.10.Hk,75.40.Cx}

\maketitle

With the extensive degeneracy of its classical ground-state, the antiferromagnetic Heisenberg model on the kagome lattice was  recognized early on as the paradigm of a quantum spin liquid phase.~\cite{Balents10}
The recent experimental discovery of  Herbersmithite, a kagome compound fluctuating down to temperatures thousands of  times lower than the coupling constant,  strengthens this speculation.~\cite{PhysicsToday_spinliquid}
Despite  numerous efforts, the nature of the ground state (GS) of the spin-$1/2$ Heisenberg antiferromagnetic kagome model (AFKM) remains controversial.
Exact quantum approaches point to the absence of long range order.~\cite{LeungElser93,Kag_ED,DMRG_kag}
Although exact diagonalizations (ED) on small samples (up to N=36) leave open the question of the criticality~\cite{ED_size_effect,PRL_Hermele},
Density Matrix Renormalization Group (DMRG) calculations\cite{Yan2011} support the idea of  a true   gapped  spin liquid.

Recently, a new instability of the degenerate \textit{classical} model towards a chiral phase has been discovered.~\cite{Regular_order}
In this paper, we show at a mean-field level that the hypothesis of a chiral spin liquid holds and is consistent with numerous robust results accumulated during the last twenty years, both for spin-1/2 and in the classical limit.~\cite{harris_kag_classique, Kag_class,Huse_Rutenberg, Octupolar_kag}

\begin{figure}
 \begin{center}
     \includegraphics[trim = 180 180 180 180, clip,width=.18\textwidth]{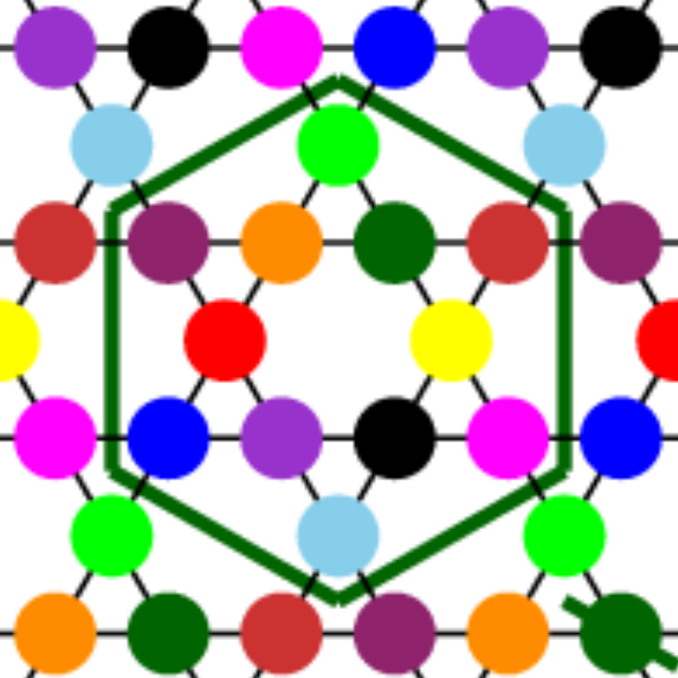}\qquad
     \includegraphics[trim = 25 140 45 150, clip,width=.19\textwidth]{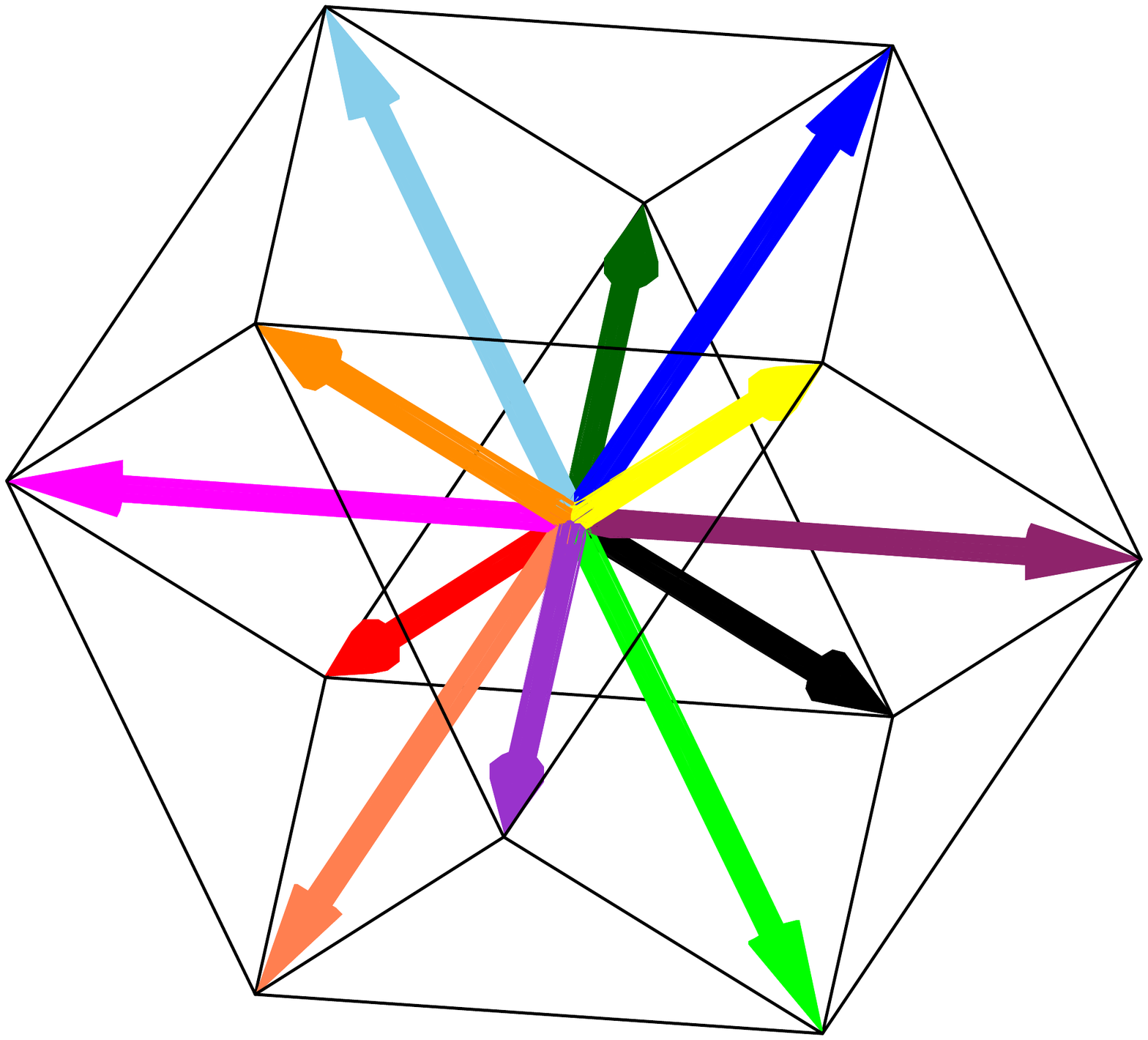}
 \end{center}
  \caption{\label{fig:cuboc1}(Color online) Description of the {\it cuboc1} order.
  Left: on the kagome lattice each color corresponds to a different magnetic sublattice.
  The thick line indicates the 12 site unit-cell.
  Right: arrows are the spin orientations with the same color coding as in the left figure.
  The black lines connecting the end of the vectors form a cuboctaedron.
  On each triangle, the spins are coplanar at 120 degrees;  for opposite sites on each hexagon, spins are anti-parallel. The triple products (determinant) of three spins of the hexagons, either first or second neighbors, are non zero and measure the chirality of the phase. They change sign in a mirror symmetry or in a spin flip.
  }
\end{figure}

The properties of chiral spin states, with simultaneously and spontaneously broken space reflection (P) and time reversal (T) symmetries, were largely debated at the end of the eighties in the wake of the Quantum Hall effect. A revival of these topics has occured thanks to graphene and flat band insulators.
\textcite{Wen_Wilczek} defined the chiral phases through the fluxes of the underlying gauge fields and \textcite{Kalmeyer_89} proposed to describe spin liquids by Laughlin wave functions. \textcite{Yang_93} suggested that the Heisenberg model on the kagome lattice might be in a chiral spin liquid state.
We reexamine this suggestion inspired by the knowledge of the classical non-planar spin order, described in Fig.~\ref{fig:cuboc1}, and propose a specific \textit{chiral  spin liquid} as the GS of the spin-$1/2$ AFKM.

\begin{figure}
\begin{center}
	\includegraphics[trim= 60 395 100 125,clip,width=.4\textwidth]{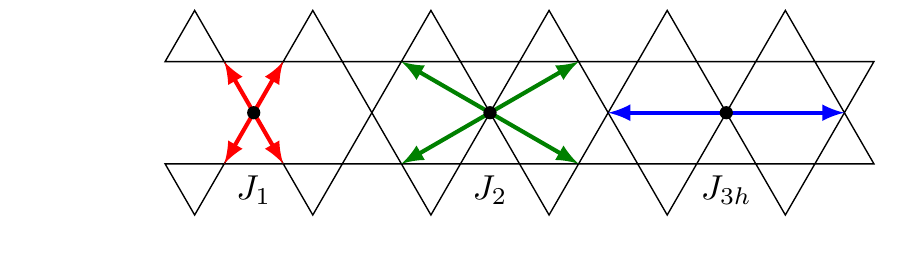}\\
	\includegraphics[trim= 2 0 0 0,clip,width=.22\textwidth]{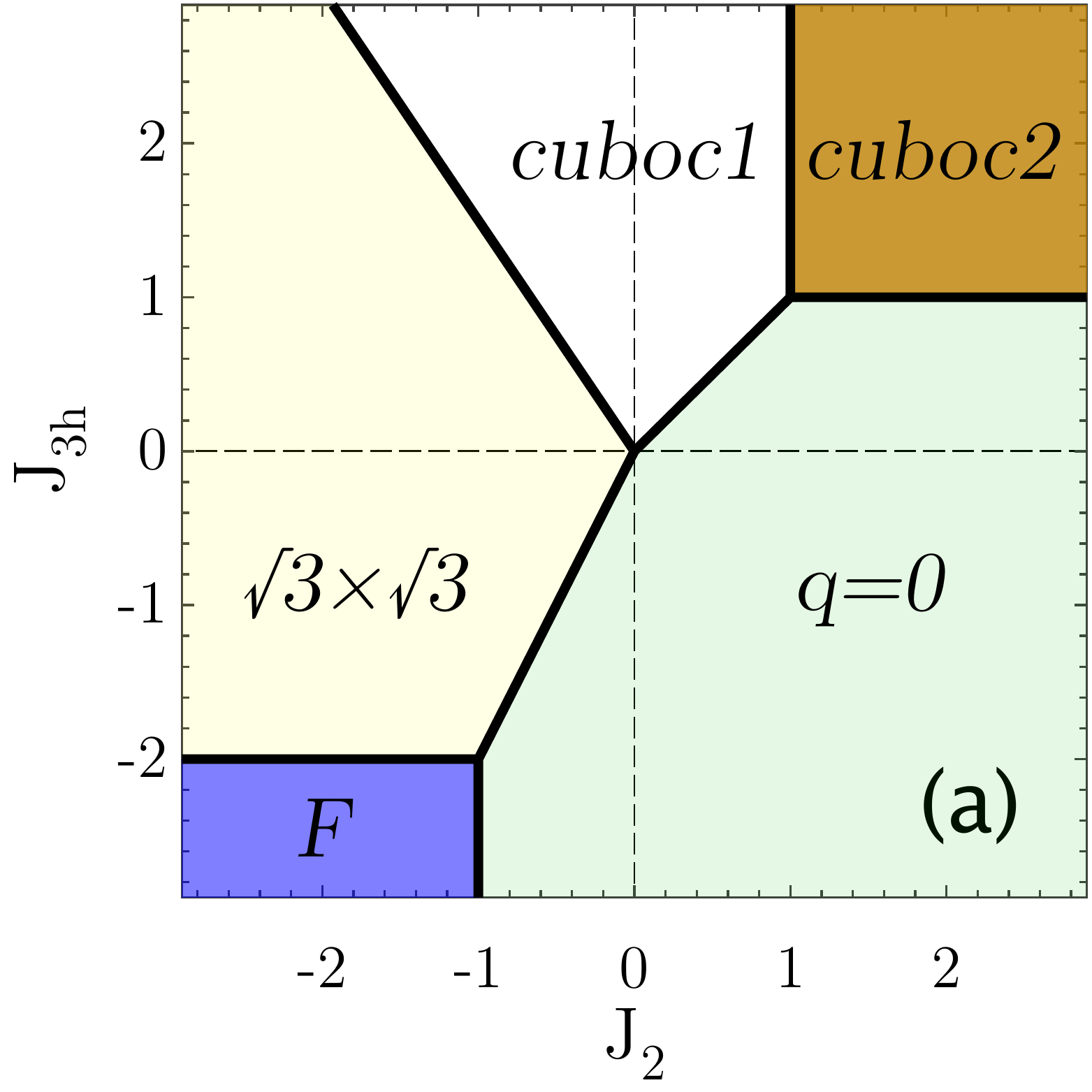}
	\includegraphics[trim= 2 0 0 0,clip,width=.22\textwidth]{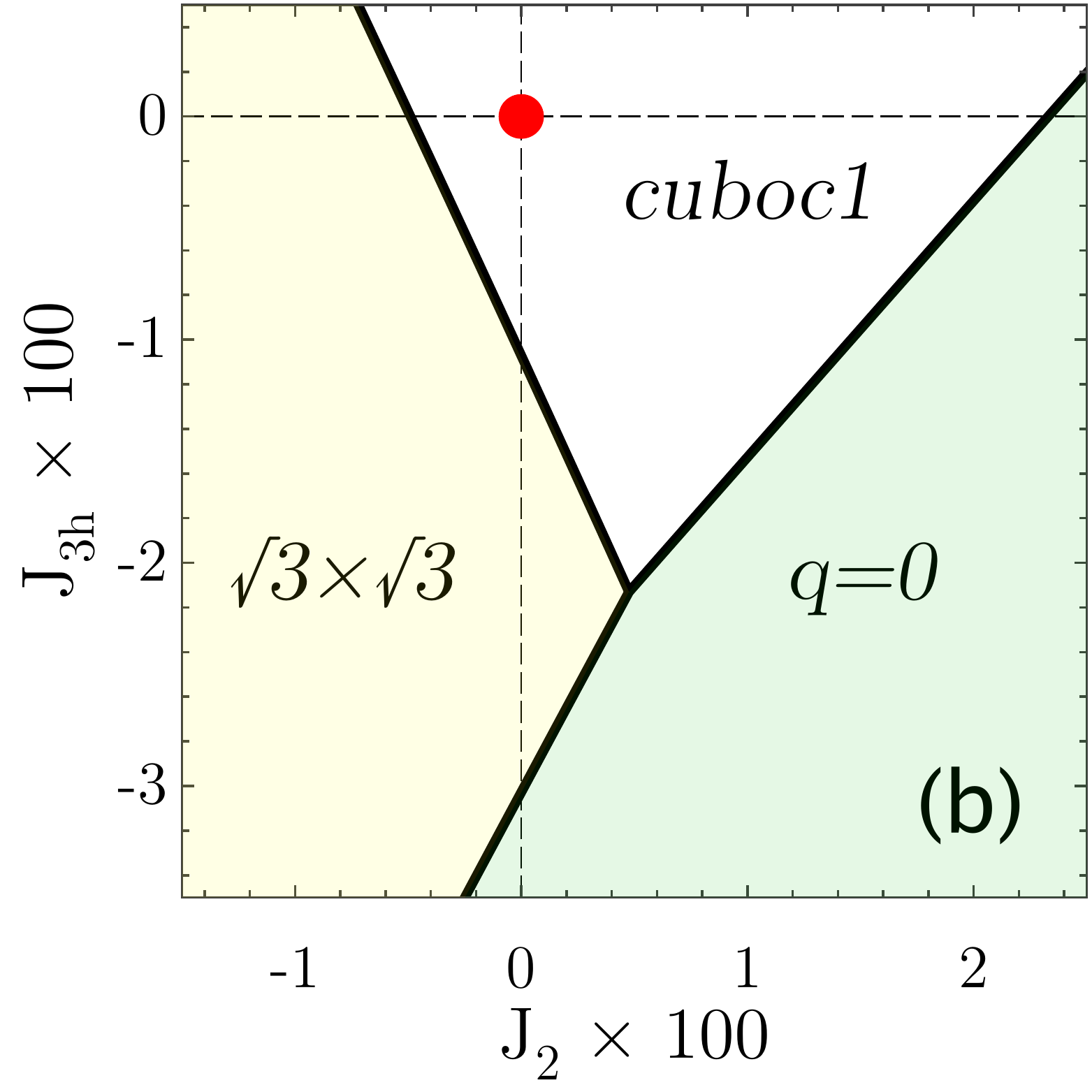}
\end{center}
\caption{
\label{fig:diag}
(Color online) Phase diagram of the model with up to third neighbors interactions at $T=0$ and $J_1=1$.
$(a)$ exact classical phase diagram (the method used to obtain this phase diagram and all orders are described in \cite{Regular_order}).
The point AFKM: $(J_2,J_{3h})=(0,0)$ is a tricritical point.
$(b)$ SBMFT phase diagram for ${\mathcal S}=0.5$.
The tricritical point stands at $(J_2,J_{3h})=(0.0049,-0.021)$ and the AFKM point (red circle) is now inside the {\it cuboc1} phase.
}
\end{figure}

The kagome lattice can be viewed as a lattice of corner sharing triangles.
The classical ground-state on a single triangle is planar with three spins at $120$ degrees.
Fixing the spin plane on a triangle does not fix the planes on adjacent ones whence the extensive ground-state degeneracy.
This degeneracy can be lifted by couplings beyond nearest neighbors.
We consider the following $J_1$-$J_2$-$J_{3h}$ Hamiltonian with $J_1=1$:
\begin{equation}
 H=\sum_{\langle i,j\rangle_\alpha} J_\alpha \widehat{\mathbf S}_i \cdot \widehat{\mathbf S}_j
\label{eq:Heisenberg}
\end{equation}
where the coupling constants and \textit{exact} classical phase diagram are given in Fig.~\ref{fig:diag}.
AFKM refers to this model when $J_2=J_{3h}=0$.
For $J_{3h}=0$, numerous  studies of quantum states have been motivated by the two planar classical states denoted \qzero ($J_2>0$) and \sqrtsqrt  ($J_2<0$)\cite{Sachdev,PSG,Kag_class,Huse_Rutenberg,rb93}.
An infinitesimal antiferromagnetic third-neighbor interaction across hexagons ($J_{3h}>0$) lifts the degeneracy of the classical AFKM to a 12-sublattice magnetic state where the spins point toward the corners of a cuboctaedron (Fig.~\ref{fig:cuboc1}), whence the name \textit{cuboc1}.
This order was first introduced by \textcite{composechiral2} who claim that this $J_{3h}>0$ interaction should be of experimental relevance.
This order is chiral: it breaks mirror symmetry. 

Monitoring the evolution under the effect of quantum fluctuations from the classical limit to the disordered spin-1/2 system  remains a challenge.
The  Schwinger boson mean-field theory (SBMFT) is an approximate but  versatile method to study, in an unified framework, both long range ordered (LRO) and gapped  spin liquid phases, from classical to the quantum limit.
In a first enlightening work, \textcite{Sachdev} showed that amongst the planar states, the \sqrtsqrt is more stable than the \qzero at the AFKM point.
In order to study the present model (Eq.~\ref{eq:Heisenberg}) around the AFKM point, we extend Sachdev's work in several directions as will be seen below.

The Schwinger boson operator $\widehat b_{i\sigma}^\dagger$ ($\sigma=\uparrow$ or $\downarrow$) creates a spin $1/2$ on lattice site $i$.
A physical spin S at site $i$ is represented by 2S bosons.
After a mean field decoupling, the Hamiltonian reads:
\begin{equation}
 H_{\rm MF}=\sum_{\langle i,j\rangle_\alpha} J_\alpha ({\mathcal B}_{ij} \widehat B^\dag_{ij}-{\mathcal A}_{ij}\widehat  A^\dag_{ij})+h.c.-\sum_i\lambda_i\widehat n_i+\epsilon_0,
\label{eq:Heisenberg_MF}
\end{equation}
where the bond operators are defined by
$2\widehat A_{ij}=\widehat b_{i\uparrow}\widehat b_{j\downarrow}-\widehat b_{i\downarrow}\widehat b_{j\uparrow}$,
$2\widehat B_{ij}=\widehat b^\dag_{i\uparrow}\widehat b_{j\uparrow}+\widehat b^\dag_{i\downarrow}\widehat b_{j\downarrow}$.
${\mathcal A}_{ij}$ and ${\mathcal B}_{ij}$ are the associated complex mean-field parameters to be determined by the self-consistency equations. The $\{{\mathcal A}_{ij},{\mathcal B}_{ij}\}$ set is called an ansatz.
$\lambda_i$ are Lagrange multipliers to constrain the mean boson number: $\langle\widehat n_i\rangle=2{\mathcal S}$ and
$\epsilon_0=\sum_{\langle i,j\rangle_\alpha} J_\alpha (|{\mathcal A}_{ij}|^2-|{\mathcal B}_{ij}|^2)+2{\mathcal S} \sum_i\lambda_i,$
where ${\mathcal S}$ is  a continuous real positive mean field parameter.

Most SBMFT studies use only one of the two types of parameters (${\mathcal A}$ or ${\mathcal B}$).
Recently, taking both fields,~ \textcite{Trumper_AB_SBMFT} found a much better description of the excitation spectrum of frustrated systems.
More specifically, the ${\mathcal A}$ fields  describe the singlet amplitudes whereas the ${\mathcal B}$ fields allow the description of boson hopping  amplitudes that are  a fundamental ingredient to describe the mixing of spin singlets and triplets on each bond, a mechanism that is central in quantum frustrated magnets.\cite{Symplectic_SBMFT}
In addition, because both ferromagnetic and antiferromagnetic interactions are treated on an equal footing, the phase diagram can be explored continuously around the origin regardless of the sign of coupling parameters.

Solving the full problem with two complex parameters per link and one real Lagrange multiplier per site is numerically too demanding for large lattices.
Looking for spin liquids or regular LRO\cite{Regular_order}, we assume $\lambda_i=\lambda$ and
 ${\mathcal A}$ and  ${\mathcal B}$ are invariant under lattice symmetries up to a local gauge transformation.
 Using projective symmetry groups (PSG)~\cite{Wen_PSG},
 \textcite{PSG} obtained four ansatzes where physical observables are invariant under lattice symmetries.
They are defined by the fluxes of the $\widehat  A$ operators on specific loops $(\phi_{\includegraphics[height=0.15cm]{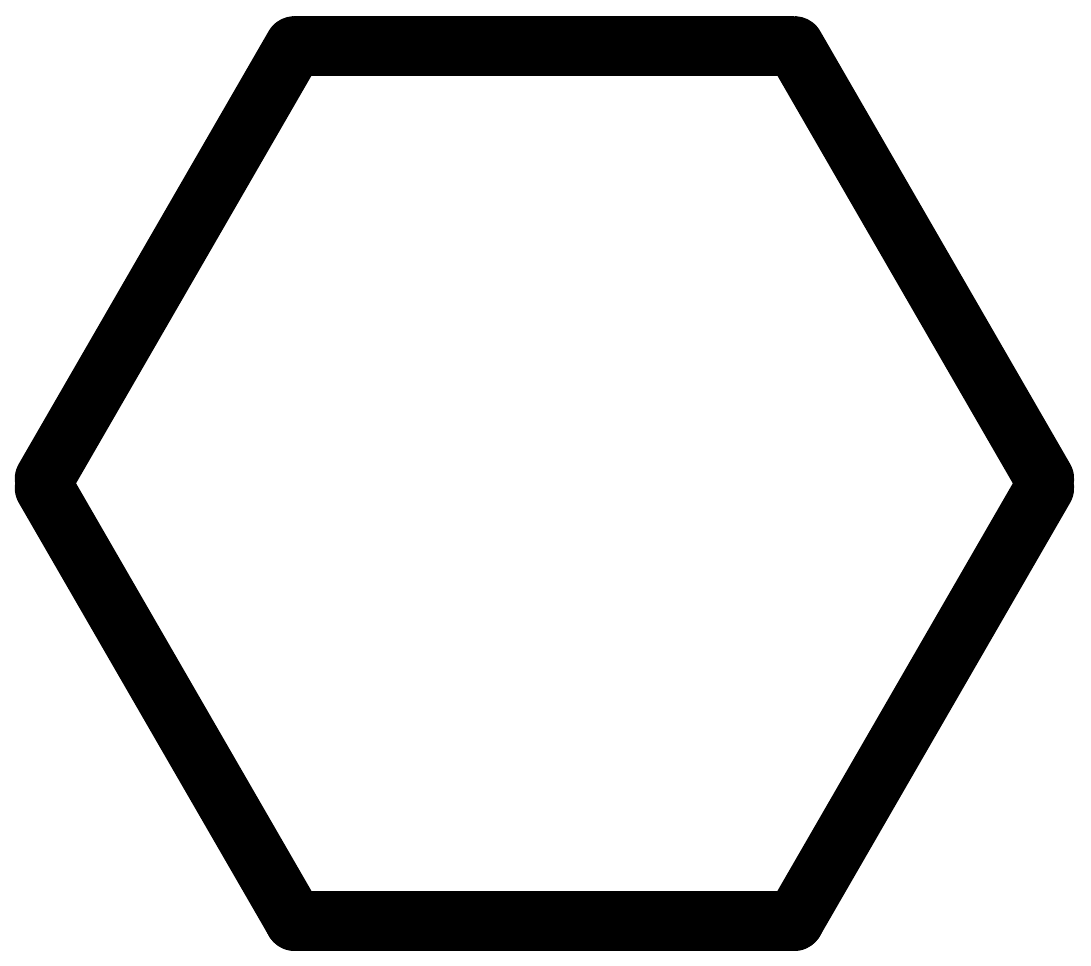}},\phi_{\includegraphics[height=0.15cm]{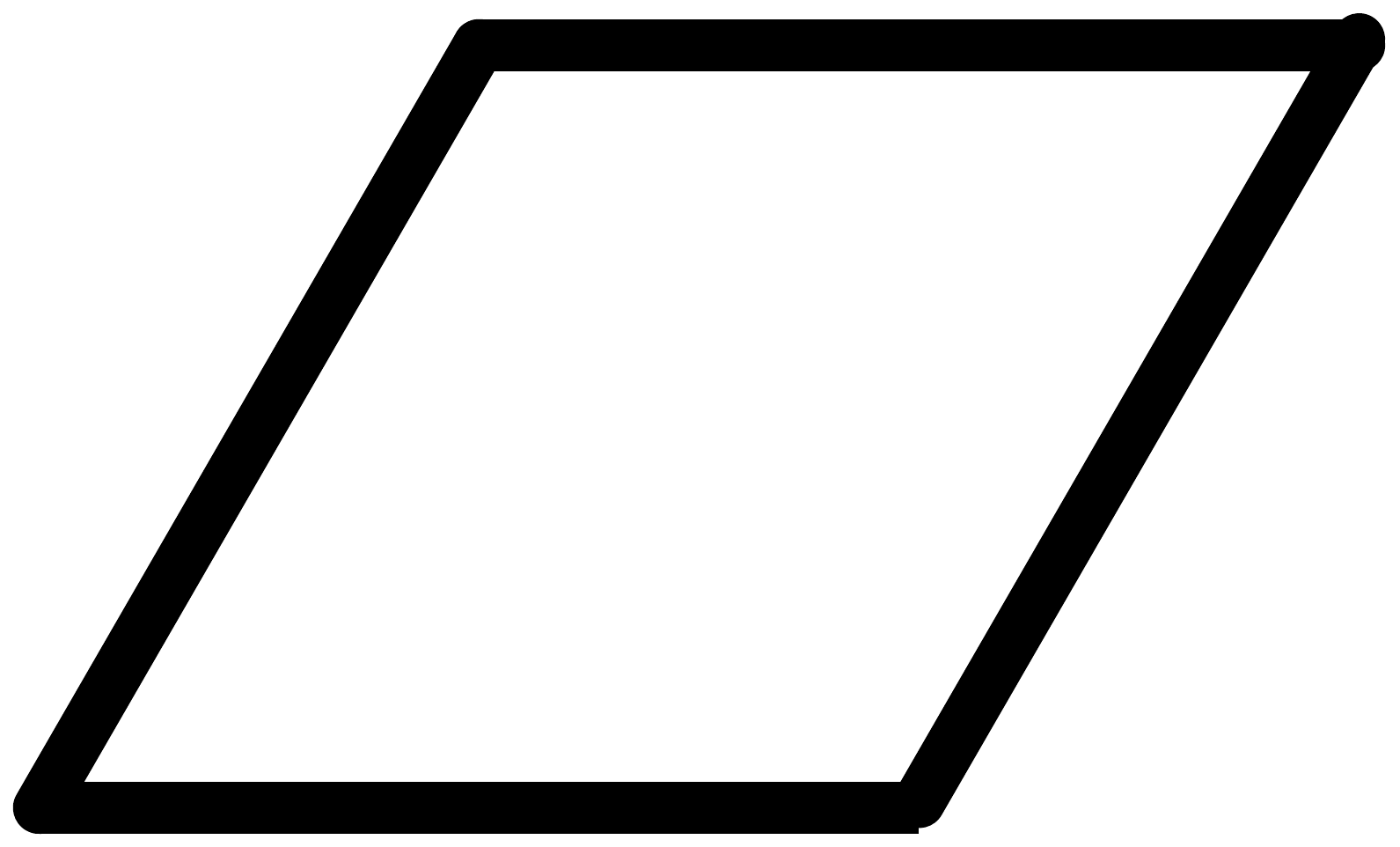}})=(0,0)$, $(\pi,0)$ , $(0,\pi)$ and $(\pi,\pi)$.
\sqrtsqrt and \qzero are associated with the first two choices respectively.

\begin{figure}
 \begin{center}
   \includegraphics[trim= 80 340 110 125,clip,width=.4\textwidth]{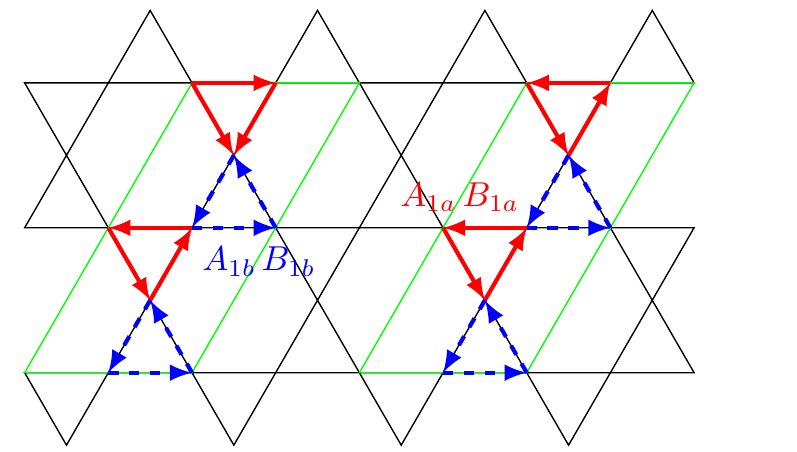}
 \end{center}
  \caption{
  \label{fig:sym_kag_Ans_cuboc1}
(Color online) Unit-cell (in thin green lines) of the {\it cuboc1} ansatz (left) and \qzero and \sqrtsqrt ansatzes (right) at the AFKM point. $A_{1a}$, $A_{1b}$, $B_{1a}$ and $B_{1b}$ are complex link parameters with constraints on their moduli: $|A_{1a}|=|A_{1b}|$, $|B_{1a}|=|B_{1b}|$ and on their arguments $\theta_{A_{1a}}=0$, $\theta_{B_{1a}}=\theta_{B_{1b}}=\pi$. The last constraint depend on the ansatz: $\theta_{A_{1b}}=0$ for \qzero, $\pi$ for \sqrtsqrt and is not fixed for {\it cuboc1}.
}
\end{figure}

The chiral {\it cuboc1} ansatz, however, cannot be obtained within this first PSG approach, because in chiral states the lattice symmetries are respected only up to a time reversal symmetry.
In all previous studies, the mean-field parameters were chosen real, the fluxes equal to $0$ or $\pi$, thus excluding  chiral ansatzes.
The extension of the PSG to include both the symmetric and the chiral spin liquids will be described in a longer paper~\cite{to_come}.
In short, the new ansatzes are defined by complex fields with specific constraints on the moduli and arguments.
Thanks to the PSG analysis the number of parameters at the AFKM point is limited to 2 moduli of bond fields for \qzero or \sqrtsqrt plus a phase $\theta_{A_{1b}}$ for \textit{cuboc1}. 
Other bond fields of the 6-spin unit cell are fixed by  algebraic constraints (see Fig.~\ref{fig:sym_kag_Ans_cuboc1} and Table \ref{tab:AB_values}). 
Non zero fluxes\cite{flux} (modulo $\pi$) induced by $\theta_{A_{1b}}$ are an indirect mean-field measure of the chirality of \textit{cuboc1}, (they are 0 or $\pi$ for \qzero or \sqrtsqrt respectively).

\begin{table}
\begin{center}
 \begin{tabular}{|c|c|c|c|c|}
\hline
                                       &         ${\mathcal S}$      &$|A_{1}|$&$|B_{1}|$&$\theta_{A_{1b}}$\\
\hline
\multirow{2}{*}{\rotatebox{0}{\qzero}}&$1/2$             &0.51624&0.18036&0\\
\cline{2-5}
                                       &$\infty$          &$\sqrt3{\mathcal S}/2$&${\mathcal S}/2$&0\\
\hline
\multirow{2}{*}{\rotatebox{0}{\sqrtsqrt}}&$1/2$             &0.51706 &0.17790&$\pi$\\
\cline{2-5}
                                       &$\infty$          &$\sqrt3{\mathcal S}/2$&${\mathcal S}/2$&$\pi$\\
\hline
\multirow{2}{*}{\rotatebox{0}{{\it cuboc1}}}&$1/2$             &0.51660 &0.17616 &1.9525 \\
\cline{2-5}
                                       &$\infty$          &$\sqrt3{\mathcal S}/2$&${\mathcal S}/2$&$1.9106^{\dag}$\\
\hline
 \end{tabular}
\caption{\label{tab:AB_values}
Values of the self-consistent SBMFT parameters for the three competing ansatzes near the AFKM point.
$ ^\dag$The exact value is $\pi-\arctan\sqrt8$.
}
\end{center}
\end{table}

The numerical solution of the mean-field equation is found from a descent method minimizing  the sum $\Sigma$ of the squares of the (free) energy derivatives with respect to the field parameters, each single energy evaluation being maximized with respect to $\lambda$ (we stop the descent when $\Sigma<10^{-8}$).
We keep the solutions with hessians of the correct sign, positive for the $\mathcal{A}$ fields and negative for the $\mathcal{B}$ if $J>0$, and the opposite if $J<0$. These requirements imply that our solutions are stable against gaussian fluctuations.

The resulting phase diagram at  $T=0$ and ${\mathcal S}=0.5$ is given in Fig.~\ref{fig:diag}-b: at the AFKM point, the {\it cuboc1} ansatz is more stable than any other regular ansatz, with an energy per site of $-0.4717.$~
\footnote{Note that SBMFT is not a variational approach, and energies may be lower than the exact one. For $N$=12, the projection of this wave-function in spin-space gives the exact energy  up to 3 digits.
Work is in progress to do this projection for larger samples.}
The numerical values of the parameters at this point are given in Tab.~\ref{tab:AB_values}.
The  parameter range  of stability  of the {\it cuboc1} phase increases when the spin decreases from $\mathcal{S}=0.5$ to $0.366$.
For $J_{3h}=0$ and $\mathcal{S}=0.5$ it is $J_2 \in [-0.005;0.025]$ and for $\mathcal{S}=0.366$ it is enlarged to $J_2 \in [-0.008; 0.045]$.
This increase is another proof of the role of quantum fluctuations in the stabilization of the {\it cuboc1} phase.

\begin{figure}
 \begin{center}
 \includegraphics[trim=11 12 13 35,clip,width=.235\textwidth]{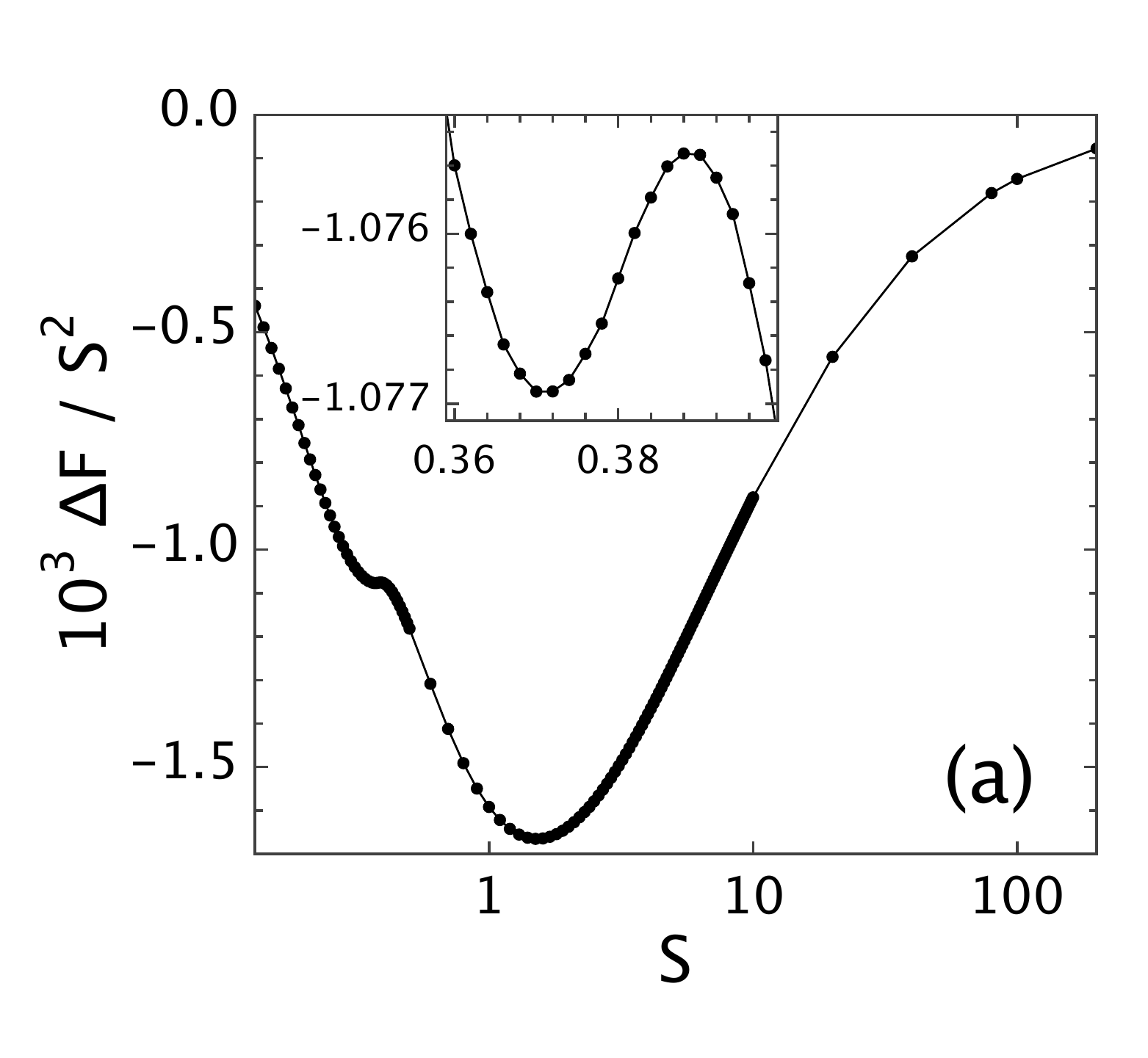}
 \includegraphics[trim=11 12 13 35,clip,width=.235\textwidth]{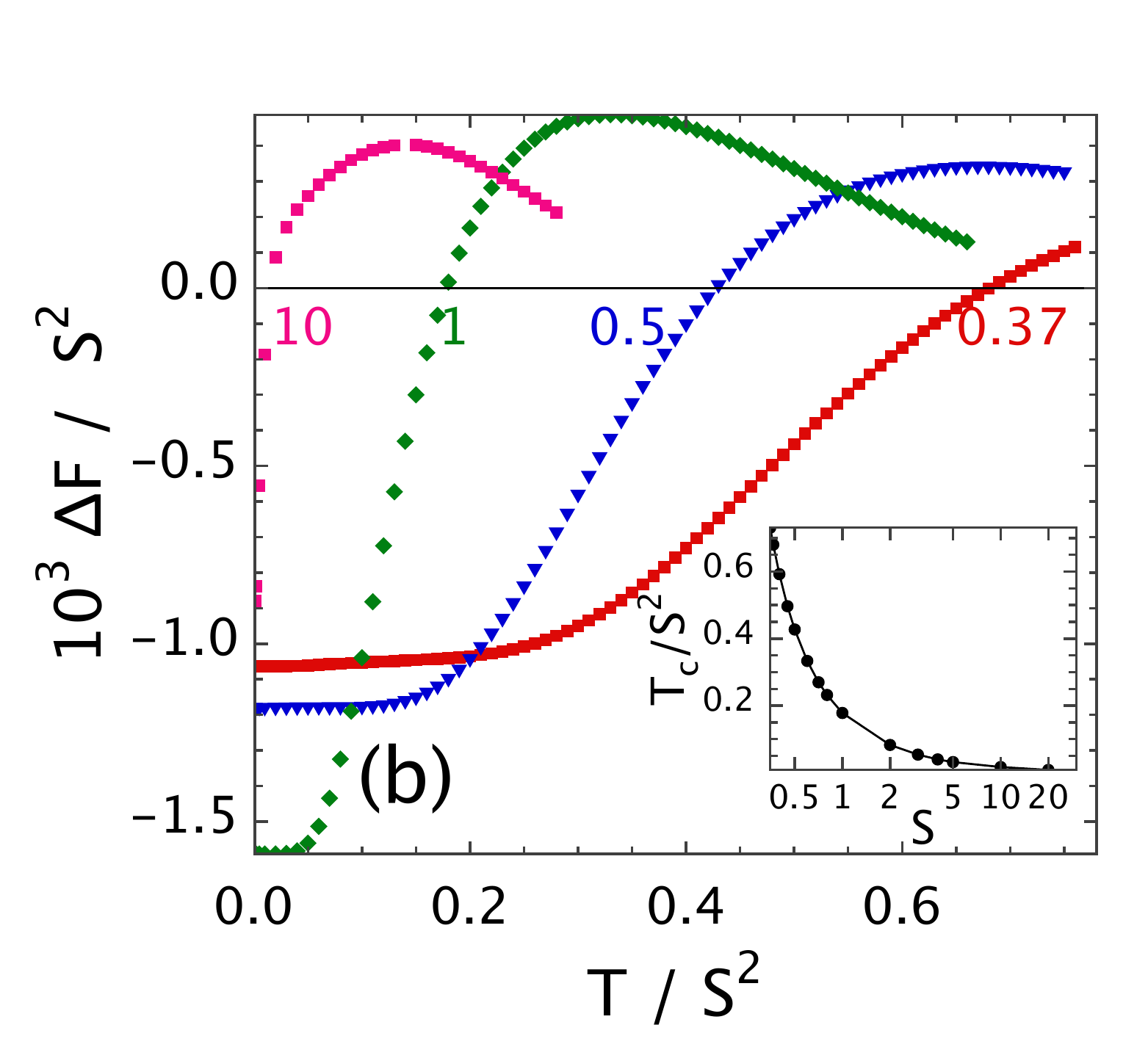}
  \caption{\label{fig:DeltaF}
    (a) Difference of SBMFT free energy between the {\it cuboc1} and \sqrtsqrt ansatzes as a function of ${\mathcal S}$ at $T=0$.
   Inset: zoom of the domain around critical spins ${\mathcal S_c}$.
    (b) Same quantity as a function of $T$ for different ${\mathcal S}$. Inset: value of $T_c/\mathcal S^2$ versus $\mathcal S$ (see the text).
    }
 \end{center}
\end{figure}

The dimensionless free energy difference ($\Delta F/\mathcal S^2$) between the \sqrtsqrt phase and the {\it cuboc1} phase is given as a function of ${\mathcal S}$ in  Fig.~\ref{fig:DeltaF}-(a).
At $T=0$, it is of the order of $10^{-3}$ in favor of {\it cuboc1}.
In the classical limit, ${\mathcal S}\to\infty$, the two phases are degenerate as they should be.
The comparison with the \qzero state is not shown as it always has  a much higher energy at the AFKM point.

Decreasing ${\mathcal S}$ leads to a second order phase transition from a gapless LRO \textit{cuboc1} phase to  a fully gapped chiral spin liquid at a critical value ${\mathcal S}_c=0.4$.
One should not hastily conclude that the true spin-1/2  system has N\'eel long range order. 
In this mean-field approach, the on-site number of bosons fluctuates: it is only fixed on average,  $\mathcal S$ is a parameter and
$\langle\widehat{\mathbf{S}}^2\rangle=3{\mathcal S}({\mathcal S}+1)/2$\cite{Auerbach}.
For spin-1/2, the good quantum number is $\langle\widehat{\mathbf{S}}^2\rangle=3/4$.
To recover this good quantum number we should use the parameter ${\mathcal S}=(\sqrt3-1)/2 \sim 0.366$.
With ${\mathcal S}=0.5$ the phase is gapless and  finite size-scaling shows a very small stiffness,
while with ${\mathcal S} \sim 0.366$ the system is  a gapped spin liquid compatible with the results of Yan \textit{et al.}.  

Now, we  turn to the effect of thermal fluctuations at the AFKM point. Fig.~\ref{fig:DeltaF}-b shows how they destabilize the {\it cuboc1} phase in favor of the \sqrtsqrt one.
The renormalized  transition temperature $T_c/{\mathcal S}^2$ from  the {\it cuboc1} to \sqrtsqrt phase decreases to zero with increasing ${\mathcal S}$ (Inset Fig.4-b).
This is consistent with classical numerical simulations showing a selection of a planar state by thermal fluctuations.\cite{Kag_class,Huse_Rutenberg,rb93,Octupolar_kag}

This chiral order hypothesis also explains why the ED spectra for sizes up to $N=36$ have a large number of singlets below the triplet gap.
Let us consider a cell of 12 spin-1/2 describing some short range order (SRO).
Assuming three spin directions (as for \sqrtsqrt and \qzero SRO), one finds a single $S$=0 ground-state derived from the coupling of three spin-2 dressed by quantum fluctuations.  With a  12-sublattice \textit{cuboc1} SRO, one finds 132  singlets built starting from the angular addition of 12 spin-1/2,  many of which are low energy states. This crude picture makes it possible to understand why there are so many (63) singlet states below the triplet gap of the $N=36$ sample. Moreover this property may explain, through resonances,  the stabilization of \textit{cuboc1} relative to \sqrtsqrt SRO.

On the other hand
with the hypothesis of a  chiral spin liquid we expect  an 8-fold GS degeneracy on a 2-torus (a factor 2 for the chirality times a factor 4 for the topological degeneracy) at the thermodynamic limit.
Thus, the large number of low lying singlets seen in ED spectra should be restricted to small size samples.
Such an evolution has already been observed in the $J_1$-$J_2$ model on the triangular lattice in the parameter range where the classical ground-state has a 4-sublattice unit cell.~\cite{lblp95}
Note that DMRG results do not exhibit a large number of singlet states below the triplet gap.\cite{Yan2011}

Spin-1/2 ED results on the $N=36$ sample are also compatible with {\it cuboc1} short range order:
$i)$ the first $S=1$-state is at the softest $k$-vector of the {\it cuboc1} short range order as can be seen in Fig.4 of ref.~\cite{ED_size_effect};
$ii)$ the dynamical and static structure factors\cite{Laeuchli} have relatively larger values at the wave vector of the {\it cuboc1} order than near the quasi soft points of the \sqrtsqrt  and \qzero  orders.
Thus, correlation functions in large scale DMRG computations and/or characterization of low energy excitations by ED for 48 sites samples would be an essential complement to further support or discard the present proposal.

Moreover, the first spin-$1/2$ states of small samples (with an odd-integer number of sites) have non-zero Chern numbers.\cite{Kag_ED}
This quantum number,
first  introduced in such a context by \textcite{Chiral_Chern},
is a topological index (and thus a robust property) characterizing the chiral character of a  wave function.
This property, which has never been explained in other approaches, could be understood for chiral spinons.\cite{Kalmeyer_89}
In the same spirit, the classical chirality defined by the determinant of three spins (non zero on hexagons for the classical \textit{cuboc1}) is generalized for quantum spins as the imaginary part of the cyclic permutation operator of spins on  closed contours.\cite{Wen_Wilczek} This quantity  (Wilson loop operator) computed in the ED-GS for different contours obeys the law expected for a chiral liquid~\cite{to_come}.

We have shown, that within mean field Schwinger boson approximation, the kagome antiferromagnet has a chiral ground state.
Spin-1/2 exact results both from ED and DMRG give some support to this hypothesis.
In such a system, we expect a low temperature chiral symmetry breaking phase with topologically protected edge states, as in the quantum Hall systems.
The extent of this  low temperature phase, the nature of the phase transition, and the role of defects,- questions already addressed in classical systems\cite{MessioDomenge,Triedres}- remain open questions.

We thank Gr\'egoire Misguich and Fr\'ed\'eric Mila for many stimulating discussions. 
We acknowledge Julian Talbot for a critical reading of the manuscript.

\bibliographystyle{apsrev4-1}

%

\end{document}